\documentclass[12pt]{article}
\usepackage[dvips]{graphicx}
\usepackage{amsmath}
\usepackage{amsfonts}
\usepackage{subfig}

\begin{document}

\renewcommand{\vec}[1]{ \boldsymbol{#1} }

\title{New physical wavelet 'Gaussian Wave Packet'}
\author{Maria V Perel and Mikhail S Sidorenko}

\maketitle \noindent
{\footnotesize\center  Department of Mathematical Physics, Physics Faculty, \\
St.Petersburg University, \\
Ulyanovskaya 1-1, Petrodvorets, St.Petersburg, 198904, Russia \\
\texttt{{mailto: perel@mph.phys.spbu.ru, M.Sidorenko@ms8466.spb.edu}
} \\}

\indent
\begin{abstract}
An exact solution of the homogeneous  wave equation, which was found
previously,   is treated from the point of view of  continuous
wavelet analysis (CWA). If time is a fixed parameter, the solution
represents a new multidimensional mother wavelet for the CWA. Both
the wavelet and its Fourier transform are given by explicit formulas
and are exponentially localized.  The wavelet is directional. The
widths of the wavelet and  the uncertainty relation are investigated
numerically. If a certain parameter is large, the wavelet behaves
asymptotically as the Morlet wavelet. The solution is a new physical
wavelet in the definition of Kaiser, it may be interpreted as a sum
of two parts: an advanced and a retarded part, both being  fields of
a pulsed point source moving at a speed of wave propagation along a
straight line in complex space-time.
\end{abstract}
\vspace{20pt}

\noindent{\it Keywords\/}: physical wavelet, acoustic wavelet,
localized wave, pulse, wave equation, mother wavelet, continuous
wavelet analysis

\newpage
\section{Introduction}

\indent

The wavelet analysis has a great many  different applications in
signal and image processing  (see \cite{Daubechies}, \cite{Mallat}),
in physics and astronomy (see \cite{Antoine}, \cite{Antoine-Book},
\cite{Berg}). It is also used for developing efficient numerical
algorithms for solving differential equations \cite{Dahmen1},
\cite{Dahmen2}; however,  mother wavelets are usually  not
associated with the solutions of differential equations under
consideration.

There is also an analytic approach to the problems of  wave
propagation proposed by Kaiser \cite{Kaiser}, where the technique of
wavelet analysis is developed for the decomposition of  solutions of
the wave equation in terms of  localized solutions, which are called
physical wavelets. They are constructed by means of a special
technique of  analytic continuation of  fundamental solutions in
complex space-time and can be split into two parts: an advanced
fundamental solution and a retarded one. The physical wavelet as a
localized solution of the wave equation has also been  given  in
\cite{Visser}. Applications of  physical wavelets were discussed in
\cite{Kaiser2}-\cite{Deschenes}.

In the present paper, we treat a new wavelet, which is at the same
time a localized solution  of the homogeneous wave equation in two
or more dimensions. This solution has been previously found and
discussed in \cite{Kiselev-Perel}, \cite{Kiselev-Perel-JMP} and
generalized in \cite{Perel-Fialkovsky}. It was named the Gaussian
Wave Packet. We study its properties from two points of view. First,
the solution can be taken as a mother wavelet for  continuous
wavelet analysis if time is a parameter and can be used in  signal
processing without being connected with any differential equation.
Secondly, this solution should be regarded  as a physical wavelet,
i.e., it is an analytic continuation to the complex space-time of
the sum of  advanced and retarded parts of the field of a point
source moving at a speed of wave propagation along a straight line
and emitting a pulse that is localized in time. It is natural to
decompose  nonstationary wave fields in terms of these solutions,
using the techniques of  wavelet analysis.

The aim of the paper is a detailed investigation of  wavelet
properties of the Gaussian Wave Packet for a fixed time and its
properties as a solution of the wave equation in view of its further
application to  problems of wave propagation. For example, the
decomposition of the solution of the initial value problem for the
wave equation in terms of wavelets has been proposed by us in
\cite{Perel}.

In Section \ref{sec-wavan}, we give a brief review of the main facts
of  continuous wavelet analysis in one and two dimensions.

In  Section \ref{sec-twow}, we show that the Gaussian Wave Packet
for a fixed time can be regarded as a wavelet,  give some estimates
of it, and present its Fourier transform.  We show that both the
wavelet and its Fourier transform have an exponential decay at
infinity. The wavelet has not only  zero mean but all  zero moments
as well.

In Section \ref{sec-simplas}, we discuss the asymptotic behavior of
the Gaussian Wave Packet as some of the free parameters become
large. We compare the packet with the nonstationary  Gaussian Beam
\cite{Brittingham}, i.e., with the solution of the wave equation
localized near the axis.  We give the Gaussian asymptotic of it
reducing it to the  Morlet well-known wavelet \cite{Antoine-Book}.

In Section \ref{sec-uncert}, we discuss the results of  numerical
calculations of the centers and widths of the packet in both the
space and spatial frequency domains. We  specify how fast these
characteristics tend to asymptotic ones, with respect to the Morlet
wavelet. We investigate  the Heisenberg uncertainty relation for
this wavelet, depending on the parameters and check how far from the
saturation it is. We also obtain results for the nonasymptotic case
where the wavelet corresponds to the solution that describes the
propagation of the wave packet of one oscillation. This case may
find applications in optics. We specify when the wavelet is
directional \cite{Antoine-article} and calculate its scale and
angular resolving powers.

Section \ref{sec-manydim} gives a generalization of the above
results to the case of an arbitrary number of  spatial dimensions.

In  Section \ref{sec-pulssour}, we establish an analogy between the
new wavelet and physical wavelets of Kaiser \cite{Kaiser}. We show
that it  may  split into incoming and outgoing parts,  each solving
a nonhomogeneous wave equation. As a source, we take  new
one-dimensional time-dependent wavelets, moving in the complex
space-time at a speed of light.

\section{Main formulas of continuous wavelet \\ analysis\label{sec-wavan}}

Wavelet analysis is a method for analyzing  local spectral
properties of functions  (for example, see \cite{Daubechies} -
\cite{Antoine}). Wavelet analysis also allows one to represent any
function of finite energy as the superposition of a family of
functions called wavelets derived from one function called a mother
wavelet by shifting and scaling its argument in the one-dimensional
case and also by rotating it in the case of several spatial
dimensions. By analogy, the Fourier transform represents a signal as
the superposition of oscillating exponents derived from one exponent
by changing its frequency.

\subsection{One-dimensional wavelet analysis}

Let us give a brief review of some basic facts concerning the
wavelet analysis of  functions dependent on one variable $x$ (for
more detail, see \cite{Daubechies} - \cite{Antoine}, \cite{Kaiser}).
Let a function $\varphi(x)$ have a zero mean, and let it decrease as
$|x|$ tends to infinity so fast that $\varphi(x) \in
\mathbb{L}_1(\mathbb{R})\bigcap\mathbb{L}_2(\mathbb{R})$. It must
oscillate to be nonzero and to have the zero mean. We call such a
function a 'mother wavelet' because we derive a two-parametric
family of functions from it, using two operations that shift the
argument by $b$ and scale it by $a$:
\begin{equation}\label{W-family-1d}
\varphi^{a,b}(x)=\frac{1}{|a|^{1/2}} \,
\varphi\left(\frac{x-b}{a}\right), \qquad b\in(-\infty, \infty),
\,\, a\in(-\infty, +\infty).
\end{equation}
Thus any function $\varphi^{a,b}(x)$ from this family again has the
shape of $\varphi$, but shifted and dilated. By means of these
operations, we can 'place' $\varphi(x)$  at any point of the $x$
axis and change its 'size' to any size by the parameter $a$. Then we
define the wavelet transform $W(a,b)$ of any signal $f(x)$ by the
formula
\begin{equation}\label{Transform-1d}
W(a,b) = \int\limits_{\mathbb{R}} \,   d  x' \, f(x') \,
\overline{\varphi^{a,b}(x')},
\end{equation}
where the bar over $\varphi$ denotes  complex conjugation.

One of the best-known mother wavelets is the Morlet wavelet, which
is
\begin{equation}\label{Morlet-1d}
\varphi(x) = \exp\left(- \frac{x^2}{2\sigma^2}\right) \left[\exp(- i
\kappa x) - \exp(-\kappa^2 \sigma^2/2) \right].
\end{equation}
It is the difference of a Gaussian function, filled with
oscillations, and a term that  provides the zero mean of
$\varphi(x)$ and that is negligible if $\kappa \sigma \gg 1.$ It is
clear from formulas (\ref{W-family-1d}) - (\ref{Morlet-1d}) that
$|W(a,b)|^2$, defined by (\ref{Transform-1d}), provides information
about the frequency content of the signal $f$ in the vicinity of the
size $a\sigma$ of the point $b$, and $\kappa/a$ plays the role of a
spatial frequency. So we may regard the wavelet transform as a
window transform, the size of a window changing for different
frequencies. Changing the size of the window makes the wavelet
transform more precise as compared to the Window Fourier (or Gabor)
transform.

We can also reconstruct the signal $f(x)$ from its wavelet transform
$W(a,b)$, or, in other words, represent the signal $f(x)$ as a
superposition of elementary signals $\varphi^{a,b}(x)$. Moreover,
the mother wavelet used for the reconstruction of $f(x)$ may differ
from the one used for the analysis.  The reconstruction formula
looks like this
\begin{equation}\label{Reconstr-1d}
f(x) = \frac{1}{C_{\varphi\psi}} \int\limits_{0}^{\infty} \, \frac{
 d  a}{a^2} \, \int\limits_{-\infty}^{\infty} \,  d  b \, W(a,b)
\, \psi^{a,b}(x),
\end{equation}
where $\psi(x)$ is another mother wavelet, and the constant
$C_{\varphi\psi}$ reads
\begin{equation}\label{C-phi-psi}
C_{\varphi\psi} = \int\limits_{-\infty}^{\infty} \,   d  k \,
\frac{\widehat{\varphi}(k) \overline{\widehat{\psi}(k)}}{|k|},
\end{equation}
where the symbol  $\widehat{}$  denotes the Fourier transform. If we
use the same wavelet for the transform and reconstruction, we should
put $\widehat{\psi}(k) = \widehat{\varphi}(k)$ and get
$|\widehat{\varphi}(k)|^2$ in this formula to calculate the
coefficient $C_{\varphi\varphi}\equiv C_{\varphi}.$

\subsection{Wavelet analysis in two  dimensions}

Wavelet analysis  can also be defined for the case of more than one
dimension (see \cite{Antoine-Book}, \cite{Murenzi},
\cite{Torresani}). A mother wavelet in the case of two dimensions
$\vec{ r} = (x, y)$ is a function $\varphi(\vec{ r})\in
\mathbb{L}_1(\mathbb{R}^2)\bigcap\mathbb{L}_2(\mathbb{R}^2)$ that
has  zero mean. The  Morlet wavelet in two dimensions reads
\begin{equation}\label{Morlet-2d}
\varphi(\vec{r}) = \exp\left(-\frac{x^2}{2\sigma_\mathrm{x}^2}-
\frac{y^2}{2\sigma_\mathrm{y}^2} \right) \left[\exp(- i \kappa x) -
\exp(-\kappa^2 \sigma_\mathrm{x}^2/2) \right].
\end{equation}
We define a family of wavelets from the mother wavelet, introducing
rotations as well as dilations and the vector translations as
follows:
\begin{equation}\label{W-family-2d}
\varphi^{a, \alpha, \vec{ b}}(\vec{ r}) = \frac{1}{a}
\varphi\left(\vec{ M}_{\alpha}^{-1} \frac{\vec{ r}-\vec{ b}}{a}
\right), \qquad \vec{ M}_{\alpha}^{-1} = \left(\begin{array}{cc}
\cos\alpha & -\sin\alpha \\
\sin\alpha & \cos\alpha
\end{array}\right).
\end{equation}
The wavelet transform is defined as
\begin{equation}\label{Transform-2d}
W(a, \alpha, \vec{ b}) = \int\limits_{\mathbb{R}^2} \,   d ^2\vec{
r'} \, f(\vec{ r'}) \, \overline{\varphi^{a,\alpha, \vec{ b}}(\vec{
r'})}, \quad \vec{ r'} = (x', y'), \ \    d ^2\vec{ r'} =  d  x' \,
 d  y'.
\end{equation}
Then the reconstruction formula takes the form
\begin{equation}\label{Reconstr-2d}
f(\vec{r}) = \frac{1}{C_{\varphi\psi}} \int\limits_{0}^{\infty} \,
\frac{ d  a}{a^3} \, \int\limits_{\mathbb{R}^2} \,   d ^2\vec{b} \,
\int\limits_{0}^{2\pi} \,  d \alpha \, W(a,\alpha,\vec{ b}) \,
\psi^{a,\alpha,\vec{ b}}(\vec{ r}),
\end{equation}
where
\begin{equation}\label{C-phi-psi-2d}
C_{\varphi\psi} = \int\limits_{\mathbb{R}^2} \,   d ^2\vec{k} \,
\frac{\widehat{\varphi}(\vec{k}) \overline{\widehat{\psi}(\vec{
k})}}{|\vec{k}|^2},\qquad \vec{k} = (k_\mathrm{x}, k_\mathrm{y}),
\qquad   d ^2\vec{ k} =  d  k_\mathrm{x} \,  d  k_\mathrm{y},
\end{equation}
and the Fourier transform $\widehat{\psi}(\vec{k})$ is
\begin{equation}
\widehat{\psi} (\vec{ k}) = \int\limits_{\mathbb{R}^2}  d ^2\vec{r}
\,\psi(\vec{r})\, \exp{(- i \vec{k} \cdot \vec{r})}, \qquad
 d ^2\vec{ r}=  d  x \,  d  y.
\end{equation}

\section{New two-dimensional wavelets \label{sec-twow} }

We consider a family of functions $\psi(\vec{r})$ of two spatial
variables $x,y$ containing arbitrary real parameters $t, \nu$ and
positive parameters $p$, $\varepsilon,$ $\gamma$ :
\begin{eqnarray}
& &\psi(\vec{r}) = \sqrt{\frac{2}{\pi}}  \; \frac{(ps)^{\nu} \,
K_{\nu}(p s)} {\sqrt{x + c t -   i \varepsilon} }, \qquad \vec{
r}=(x,y),
\label{packet3} \\
& & s =  \sqrt{1 -   i {\theta}/{\gamma}}, \qquad \qquad \label{def-s} \\
& & \theta = x - ct + \frac{y^2}{x + ct -  i \varepsilon},
\label{theta}
\end{eqnarray}
where $K_{\nu}$ is the  Bessel modified function (MacDonald's
function) \cite{Abramovitz-Stigan}. The branch of the square root in
formula (\ref{def-s}) with positive real part is taken. The choice
of the branch of the square root in the denominator of
(\ref{packet3}) is not important, for the sake of definiteness we
assume that it has positive real part. We intend to show that each
of the functions from the family (\ref{packet3}) is suited for the
role of a mother wavelet with good properties. The same is valid for
their derivatives of any order with respect to spatial coordinates
and time.

Function (\ref{packet3}) has appeared in \cite{Perel-Fialkovsky} in
connection with  the linear wave equation
\begin{equation}
\psi_{tt} - c^2 ( \psi_{xx} + \psi_{yy} ) = 0, \qquad c=const.
\label{wave}
\end{equation}
If we regard the parameter $t$  as time, formula (\ref{packet3})
gives an exact solution of (\ref{wave}), which is well localized if
$p \gg 1$ (see below). If $\nu={1 \over 2},$ formula (\ref{packet3})
yields
\begin{equation}
\psi(\vec{r})=\frac{\exp{( - ps)}}{ \sqrt{x + ct -
 i \varepsilon}}. \label{packet}
\end{equation}
If in addition  $\varepsilon=\gamma,$ it leads to the exact solution
of (\ref{wave}) which was first reported in \cite{Kiselev-Perel} and
discussed in detail in \cite{Kiselev-Perel-JMP}.

In this section we view $\psi$ as a two-dimensional mother wavelet
with $t$ being a parameter, ignoring that it is a solution of
(\ref{wave}). According to \cite{Daubechies}, \cite{Antoine-Book},
this is possible, provided that the following conditions are
satisfied:
\begin{equation}
\int\limits_{\mathbb{R}^2} \,   d ^2\vec{r}\, |\psi(\vec{r})| <
\infty, \qquad\int\limits_{\mathbb{R}^2} \,  d ^2\vec{r}\, |
\psi(\vec{r})|^2 < \infty,\qquad   d ^2\vec{r} \equiv   d  x\,  d y,
\label{LL}
\end{equation}
i.e., $\psi(\vec{r}) \in \mathbb{L}_1(\mathbb{R}^2) \bigcap
\mathbb{L}_2(\mathbb{R}^2),$ and it has zero mean, i.e.,
\begin{equation}
\int\limits_{\mathbb{R}^2} \,   d ^2\vec{r}\, \psi(\vec{r}) = 0.
\label{mean}
\end{equation}
To prove (\ref{LL}) we note that   formulas (\ref{def-s}),
(\ref{theta}) imply that $\mathrm{Re}(s^2)\ge 1$ and thus we get
$\mathrm{Re}(s)\ge 1,$ $|\arg(s)|< \pi/4$. Therefore,
(\ref{packet3}) has neither singularities nor branch points for real
$x, y$ and $t$. It is a smooth function of $x,y,t,$ and its
derivatives of any order with respect to $x, y, t$ are also smooth
functions. We also obtain
\begin{equation}
|s^{2}|^2  =   1  + \frac{ (x - ct)^2 + y^2 } {\gamma^2 } +
\frac{y^2 ( (x-ct)^2 + y^2 + 2\varepsilon\gamma - \varepsilon^2
-4(ct)^2 ) } { \gamma^2 [ ( x + ct)^2 + \varepsilon^2 ] }.
\label{s2}
\end{equation}
If $x$ and $y$ are large to an extent that the third term in
(\ref{s2}) is positive, then
\begin{equation}
|s^{2}|^2  > \frac{ (x - ct)^2 + y^2 } {\gamma^2 }.
\end{equation}
Hence for large $x$ and $y$ the  Bessel modified function can be
replaced by its asymptotics, resulting in
\begin{equation}
\psi(\vec{ r}) \, =  \,\frac{(ps)^{\nu -1/2}\, \exp{( - p s )} } {
\sqrt{x+ct -   i \varepsilon}}\; \left[ 1 +   O  \left(\frac{1}{| p
s |} \right) \right]. \label{as-pack}
\end{equation}
Noting that $\mathrm{Re}(s)\ge |s|/\sqrt{2}$ because
$|\arg(s)|<\pi/4,$  we conclude that $\psi(\vec{ r})$ has an
exponential falloff and  $\psi \in \mathbb{L}_1 \bigcap
\mathbb{L}_2.$

To check that the condition (\ref{mean}) is satisfied, we calculate
the Fourier transform of $\psi(\vec{r})$
\begin{equation}
\widehat{\psi} (\vec{k}) = \int\limits_{\mathbb{R}^2}   d ^2\vec{r}
\,\psi(\vec{r})\, \exp{(- i \vec{k} \cdot \vec{r})} , \qquad \vec{k}
= (k_\mathrm{x}, k_\mathrm{y}).
\end{equation}
The calculations yield (see Appendix 1)
\begin{equation}
\widehat{\psi} (\vec{k}) =  2\pi \,  e ^{ i  \pi/4}
\frac{p^{2\nu}}{\gamma^\nu}  \, \left[ k(k+k_\mathrm{x})^{\nu+1/2}
\right]^{-1} \,  \nonumber
\end{equation}
\begin{equation}
\times \exp\left[ - (k + k_\mathrm{x}) \frac{\gamma}{2} - (k -
k_\mathrm{x} )\frac{\varepsilon}{2} - \frac{ p^2 }{2 \gamma
(k+k_\mathrm{x})} -   i   k c t \right], \label{Four3}
\end{equation}
where $k = |\vec{ k}|$. The formula (\ref{Four3}) shows that
$\widehat{\psi}(\vec{ k})\vert_{{k}=0} = 0,$ owing the term in the
exponent containing  the denominator $k + k_\mathrm{x}$. Therefore,
the function $\psi(\vec{ r})$ has  zero mean (\ref{mean}).
Conditions (\ref{LL}), (\ref{mean}) enable $\psi(\vec{ r})$ to be a
mother wavelet.

Moreover, the following relation holds:
\begin{equation}
\left. \frac{\partial^{l+m} \left[ k_\mathrm{x}^j \;
k_\mathrm{y}^\mu \; \widehat{\psi}(\vec{ k})\right]} {\partial^l
k_\mathrm{x} \; \partial^m  k_\mathrm{y} } \right|_{{k}=0} = 0
\label{admiss-cond-moments}
\end{equation}
for any integer nonnegative $j$, $\mu$, $l$, and $m$. This
condition, the smoothness and the exponential falloff of $\psi(\vec{
r})$ mean that any derivative of $\psi$  may be viewed as a mother
wavelet and that all the moments of the wavelet $\psi$ and its
derivatives vanish, i.e.,
\begin{equation}
\int\limits_{\mathbb{R}^2}  d ^2 {\vec{r}} \, x^l y^m \, \psi(\vec{
r}) \, = \,0, \qquad \int\limits_{\mathbb{R}^2}   d ^2 {\vec{r}} \,
x^l y^m \,\frac{\partial^{j + \mu}\; \psi(\vec{ r})} {\partial^j x
\;
\partial^\mu y } \,  = \,0. \label{moments}
\end{equation}
This property indicates that such wavelets could be useful  in
singular fields \cite{Daubechies}-\cite{Antoine-Book}.

\begin{figure}[htbp]
\centering \subfloat[Plot of the real part of $\psi(\vec{r}).$]{
\includegraphics[keepaspectratio,
width=0.8\textwidth]{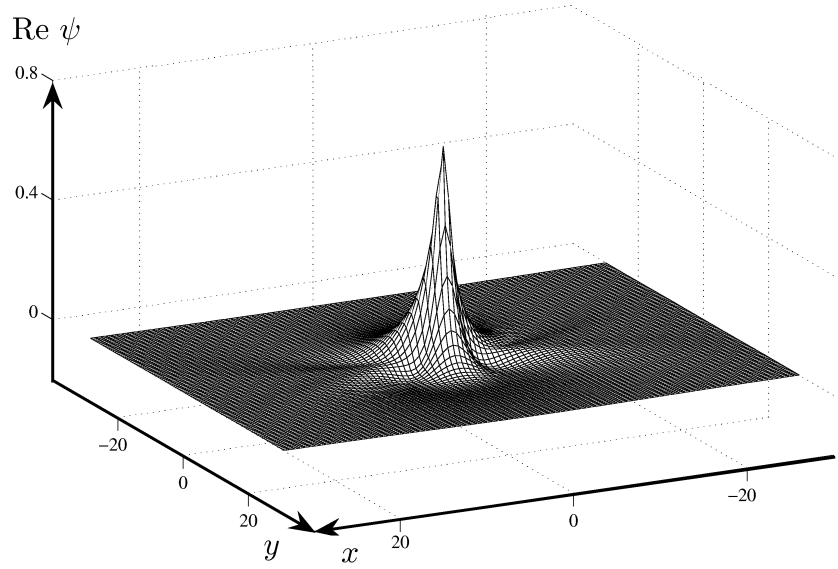} \label{pic-gaus1:cord} }  \\
\subfloat[Plot of the Fourier transform $\widehat{\psi}(\vec{k}).$]{
\includegraphics[keepaspectratio,
width=0.8\textwidth]{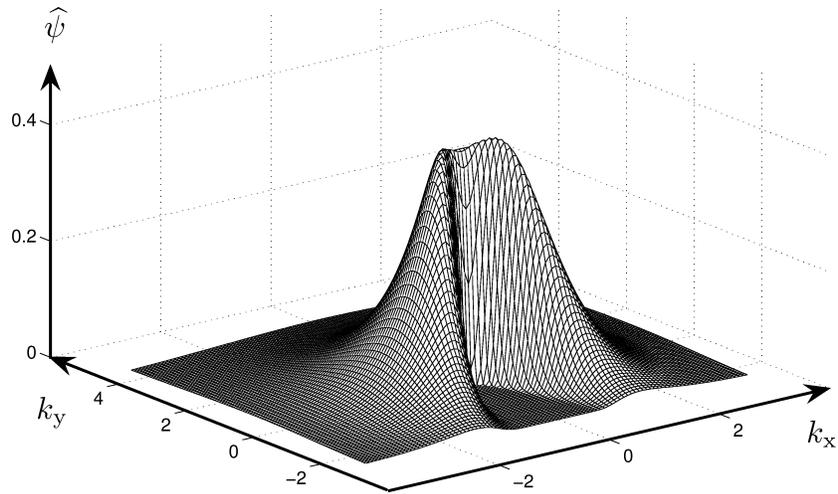} \label{pic-gaus1:four}}
\caption{An example of the new wavelet and its Fourier transform for
$p = 0.5, \,\, \varepsilon = 1, \,\, \gamma = 0.25.$ }
\label{pic-gaus1}
\end{figure}

\begin{figure}[htbp]
\centering \subfloat[Plot of the real part of $\psi(\vec{r}).$]{
\includegraphics[keepaspectratio,
width=0.8\textwidth]{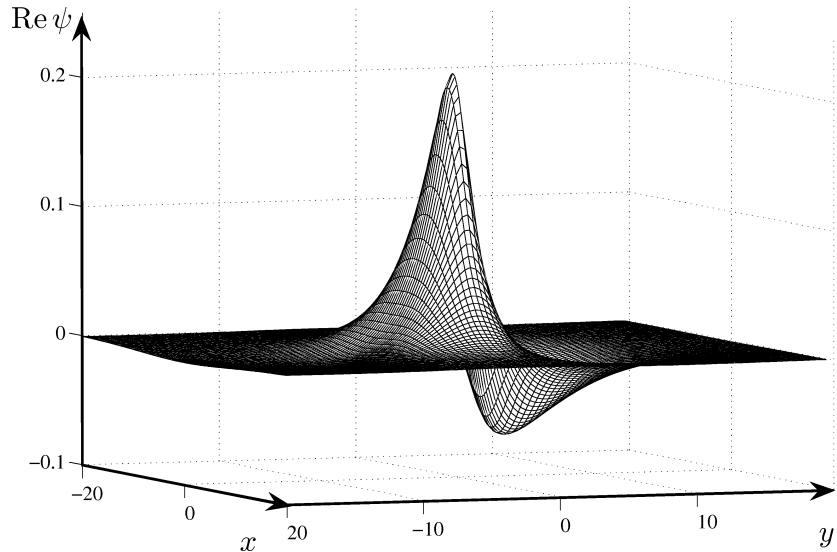} \label{pic-gaus2:cord} }  \\
\subfloat[Plot of the Fourier transform $\widehat{\psi}(\vec{k}).$]{
\includegraphics[keepaspectratio,
width=0.8\textwidth]{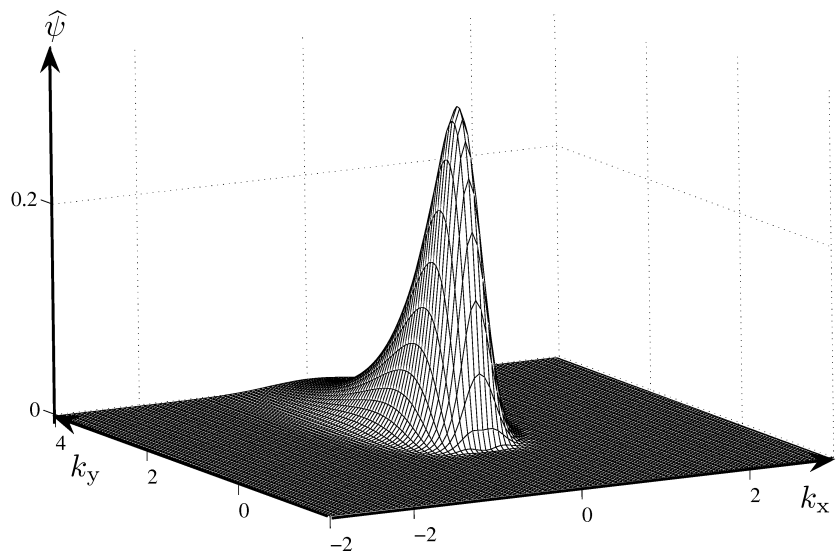} \label{pic-gaus2:four}}
\caption{An example of the new wavelet and its Fourier transform for
$p = 1, \,\, \varepsilon = 16, \,\, \gamma = 0.5.$ }
\label{pic-gaus2}
\end{figure}

The wavelet $\psi$ has simple asymptotics for large values of $p$,
which is discussed in the next section. Below we present
calculations of the wavelet (\ref{packet3}) and its Fourier
transform (\ref{Four3}) for moderate values of $p$ when no
asymptotics can be applied. It should be mentioned that the wavelet
(\ref{packet3}) represents a wave of one oscillation when $\sqrt{p}
\approx 1$ (see Figures \ref{pic-gaus1}, \ref{pic-gaus2}). This case
is applicable in optics in the case of propagation of short pulses.

\section{Simple asymptotics of the wavelet \label{sec-simplas}}

In this section, we study  asymptotic properties of the Gaussian
Wave Packet when the parameter $p$ is large. First we note that we
can replace the McDonald function in  (\ref{packet3}) by its
exponential asymptotics (see \cite{Abramovitz-Stigan}) and then we
get (\ref{as-pack}) for any $x,y,t$ if $p \gg 1$. To prove this, we
 show that $\mathrm{Re}(s) \geq 1$, then $\mathrm{Re}(ps) \geq p
\gg 1$, which provides that $|ps| \gg 1$ and the exponential
asymptotics is suitable.

We intend to show first that the modulus of the exponent in
(\ref{as-pack}) has a maximum at the point $x = ct, y = 0$. Let  $s
= a +  i  b,$ then the relation $2 a^2 = (a^2 - b^2) + (a^2 +b^2)$
yields
\begin{equation}
2 [\mathrm{Re} (s)]^2 = \mathrm{Re} (s^2) + \sqrt{ [\mathrm{Re}
(s^2)]^2 + [\mathrm{Im} (s^2)]^2}. \label{Re-s2}
\end{equation}
Hence $[\mathrm{Re} (s)]^2 \ge \mathrm{Re}(s^2)$, and taking into
account that first
\begin{equation}
\mathrm{Re} (s^2) = 1 +  \frac{\varepsilon y^2}{\gamma \;[\;(x+ct)^2
\;+\; \varepsilon^2\;]}
\end{equation}
and  secondly
\begin{equation}
|s^{2}|^2  =   1  + \frac{ (x - ct)^2  } {\gamma^2 }, \label{s3}
\end{equation}
which follows from (\ref{s2}) if $y=0$, we get $\mathrm{Re} (s) > 1$
outside the point $x = ct,$ $y=0.$ Formulas (\ref{def-s}),
(\ref{theta}) show that $s=1$ if $x = ct, y = 0,$ which is the point
of the maximum of the modulus of  exponent (\ref{as-pack}). Hence
$\mathrm{Re}(s) \geq 1$, next $\mathrm{Re}(ps) \gg 1,$ and the
asymptotics (\ref{as-pack}) can be used.

Now we  study the behavior of the Gaussian Wave Packet near its
maximum and show that it can be approximated by a nonstationary
Gaussian beam. Consider a domain near the point $x = ct, y = 0$
which is of  order
\begin{equation}
\frac{x-ct}{\gamma} =   O  \left(\frac{1}{p^{\alpha}}\right), \quad
\frac{y}{ \sqrt{\varepsilon\gamma} } =  O
\left(\frac{1}{p^{\alpha}}\right), \qquad {1 \over 3} \,<\, \alpha
\,<\, {1 \over 2}. \label{estim-vicin}
\end{equation}
We  prove below that the wavelet $\psi(\vec{ r})$ has a uniform
asymptotics in this domain as $p \to \infty$ :
\begin{equation}
\psi(\vec{r})  = \psi_{\mathrm{beam}}(\kappa,\vec{r}) \, \exp\left[
- \frac {\kappa (x - c t)^2}{4\gamma} \right] C \left[ 1 +  O
\left(p^{-3\alpha+1} \right) \right], \label{Morlet2}
\end{equation}
where $\kappa = p/(2\gamma)$ and $\psi_{\mathrm{beam}}$ is a
nonstationary Gaussian beam \cite{Brittingham}, \cite{Kiselev}
\begin{equation}
\psi_{\mathrm{beam}}(\kappa, \vec{r})  =  \frac{ \exp{(  i  \,
\kappa \, \theta)} }{\sqrt{x + ct -  i  \varepsilon}}, \label{beam}
\end{equation}
\begin{equation}
C = p^{\nu - 1/2} \exp{ (-  p) }.
\end{equation}

To prove (\ref{Morlet2}), we decompose $s$ from (\ref{def-s}) into
powers of $\theta/\gamma$, which are  of  order  $ O  (p^{-\alpha})$
by the formula (\ref{estim-vicin}) and the inequality $ \left| { {
y^2 } / {( x + ct -  i \varepsilon )} } \right| < {y^2 /
\varepsilon}$. We obtain
\begin{equation}
s = \sqrt{1 -  i {\theta}/{\gamma}} = 1 -  i {{\theta}\over{2
\gamma}} + {{\theta^2}\over{8 \gamma^2}} + {\cal E}_1.
\label{s-expand}
\end{equation}
We insert (\ref{s-expand}) into (\ref{as-pack}), substituting the
expression for $\theta$  (\ref{theta}) and the quadratic term
${{\theta^2} / {(8 \gamma^2)}} = {{ (x-ct)^2} / {(8 \gamma^2)}} +
{\cal E}_2 $ into it. It is easy to show that
\begin{equation}
\exp[-p({\cal E}_1 + {\cal E}_2) ] = 1 +  O \left(p^{- 3\alpha +
1}\right). \label{err}
\end{equation}
We also note that the multiplier $s^{\nu}$ of the McDonald function
in formula (\ref{packet3}) has an estimate $s^{\nu} = 1 +  O
(p^{-\alpha})$ and  $ O  (p^{-\alpha}) =  O (p^{-3\alpha+1})$,
because we have restricted $\alpha$ to the interval $({1 \over 3},
\, {1 \over 2}).$ The asymptotic formula (\ref{Morlet2}) is proved.

Let us discuss  formula (\ref{Morlet2}). The intervals where
$(x-ct)/{\gamma}$ and $ {y}/ \sqrt{\varepsilon\gamma}$ vary are
small compared to unit, according to (\ref{estim-vicin}). However
they are large enough to ensure the exponential falloff of the
function $\psi(\vec{ r})$ on the edges of these intervals. To
clarify this we note that the exponential terms in formulas
(\ref{Morlet2}), (\ref{beam}) are of  order  $ O (p^{-2\alpha+1}).$
They tend to infinity as $p\to\infty$ by our choice of $\alpha< {1
\over 2}$.

Asymptotics (\ref{Morlet2}) is expressed in  terms of the field of
the nonstationary Gaussian beam $\psi_{\mathrm{beam}}(\kappa,\vec{
r})$ \cite{Brittingham}. If $t$ is time,
$\psi_{\mathrm{beam}}(\kappa,\vec{ r})$ is an exact solution of the
wave equation (\ref{wave}) with infinite energy. It is localized
near the axis $ox$ because $\mathrm{Re}(  i  \kappa \theta) = -
\kappa \varepsilon y^2/ [ (x+ct)^2 + \varepsilon^2 ]$ and the cross
section of the beam attains its minimum when $x = - ct$. Formula
(\ref{Morlet2}) gives the Gaussian beam (\ref{beam}) multiplied by
the cutoff function $\exp[ - \kappa (x - c t)^2/(4\gamma) ]$. The
greater  $\gamma$, the more elongated  the essential support of
$\psi$, the closer it approximates the Gaussian beam (\ref{beam}).

Formula (\ref{Morlet2}) allows an additional simplification if
$\gamma \le \varepsilon$ and  $2ct/\varepsilon =   O (p^{-\alpha}).$
In view of estimates (\ref{estim-vicin}) and  the decomposition
\begin{equation}
\frac{p}{\gamma} \left(\frac{y^2}{x+ct-  i  \varepsilon} \right) =
\frac{p}{\gamma} \left(\frac{y^2}{-  i  \varepsilon \left[1 +  i
(x-ct)/\varepsilon + 2  i   ct/\varepsilon \right]} \right)
\nonumber
\end{equation}
\begin{equation}
=   i   p\frac{y^2}{\varepsilon\gamma} +  O
\left(\frac{\gamma}{\varepsilon} p^{-3\alpha+1} \right) +  O  \left(
p^{-3\alpha+1} \right), \label{gam-eps}
\end{equation}
the asymptotic formula (\ref{Morlet2}) takes the form
\begin{equation}
\psi(\vec{r}) =  \frac{C}{(-   i  \varepsilon)^{1/2} } \exp{\left[ i
\kappa (x - c t)  - \frac{ (x - c t)^2}{2\sigma_\mathrm{x}^2} -
\frac{y^2}{2\sigma_\mathrm{y}^2} \right] } {\left[1 +   O (
p^{-3\alpha+1}) \right]}, \label{Morlet}
\end{equation}
where
\begin{equation}\label{sigma}
\sigma_\mathrm{x}^2 = { 4\gamma^2 }/{p }, \qquad \sigma_\mathrm{y}^2
= \gamma \varepsilon/p.
\end{equation}
It is the Morlet wavelet (see \cite{Daubechies},
\cite{Antoine-Book}) with  center at the point $x = c t$, $y=0$. The
numerical essential support of the wavelet, i.e., the points in the
$(x,y)$ plane where $\psi(\vec{r})$ is not negligible numerically,
is an ellipse with  semiaxes proportional to $\sigma_\mathrm{x}$ and
$\sigma_\mathrm{y}$. The ratio $\sigma_\mathrm{x}/\sigma_\mathrm{y}
= 2\sqrt{\gamma/\varepsilon}$ rules its asymmetry. We recall that,
unlike the Morlet wavelet, the wavelet $\psi$ has  zero mean and
vanishing moments of any order.

Another asymptotic formula can be obtained from the formula
(\ref{Morlet2}) if $\gamma > \varepsilon$ and  $ct/\varepsilon > 1,$
but it is not necessarily small. This asymptotics was found and
studied in \cite{Kiselev-Perel-JMP} in detail with the restriction
$\gamma=\varepsilon$. It is equal to
\begin{equation}\label{as-pack-2}
\psi(\vec{r}, t)  \sim \exp\left[  i \kappa(x-ct) -
\frac{(x-ct)^2}{2 \sigma_\mathrm{x}^2} - \frac{y^2}{2
\widetilde{\sigma}_\mathrm{y}^2} +   i \frac{2ct\kappa
\,y^2}{4c^2t^2+\varepsilon^2}\right]\frac{C}{(2ct -   i
\varepsilon)^{1/2}},
\end{equation}
where $\sigma_\mathrm{x}^2$ is the same as in (\ref{Morlet}) and is
given in (\ref{sigma}). The width of the packet along the axis $oy$
is a function of  time $t$ and reads as
$\widetilde{\sigma}_\mathrm{y}^2 = \sigma^2_\mathrm{y}(1 + 4c^2t^2 /
\varepsilon^2).$

This asymptotics is obtained using (\ref{s-expand}) and decomposing
$1/(x+ct-  i  \varepsilon)$ inside $\theta$ in powers of $x-ct.$ The
quadratic terms in $x-ct$ and $y$ are taken into account.

\section{The uncertainty relation and  directional \\ properties \label{sec-uncert}}

In this section, we discuss  numerical properties of the Gaussian
Wave Packet,  which are important for  further applications of this
new wavelet.  We specify when the asymptotic case becomes valid. We
also consider the wavelet in a nonasymptotic situation.

\subsection{Widths of the wavelet $\psi$ and the uncertainty relation}

\begin{figure}[h]
\centering
\includegraphics[keepaspectratio,
width=0.7\textwidth]{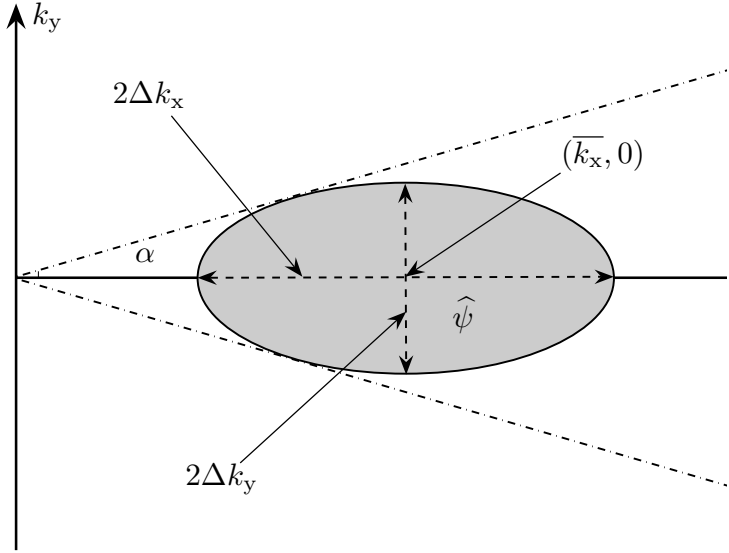}  \caption{The essential
numerical support of  $\hat{\psi}.$ } \label{pic-supp1}
\end{figure}

Let us define the centers and widths of a function $\psi(\vec{ r})$,
$\vec{ r}=(x_1, ..., x_\mathrm{n}).$ We denote by $\|\psi\|_2$ the
$\mathbb{L}_2$ norm of $\psi$:
\begin{equation}
\|\psi\|_2 = \left(\int\limits_{\mathbb{R}^n}   d ^n\vec{r} \,
|\psi(\vec{r})|^2 \right)^{1/2}, \qquad   d ^n\vec{r} =   d  x_1 \,
 d  x_2\,...   d  x_\mathrm{n}.
\end{equation}
We define the centers and widths of a function as follows
\begin{equation}\label{center}
\overline{x_i} = \frac{1}{\|\psi\|^2_2}\int\limits_{\mathbb{R}^n}
 d ^n\vec{ r} \,\,x_i\,\, |\psi (\vec{ r})|^2 ,
\end{equation}
\begin{equation}\label{width}
\Delta x_i = \frac{1}{\|\psi\|_2} \left(\int\limits_{\mathbb{R}^n}
 d ^n\vec{r} \,\,(x_i - \overline{x_i})^2 \,\, |\psi (\vec{r})|^2
\right)^{1/2}.
\end{equation}
In the two-dimensional case we put $n=2$, $x_1 = x$ and $x_2 = y$.
Using these formulas, we can also calculate the centers and widths
of the Fourier transform of a function $\psi$ by replacing
$\psi(\vec{ r})$ by $\widehat{\psi}(\vec{ k}).$ We suppose that the
essential numerical support of the wavelet, i.e., the points in the
$(x,y)$ plane where $\psi(\vec{ r})$ is not negligible numerically,
is the ellipse with semi-axes $\Delta x$ and $\Delta y$. The
essential numerical support of the Fourier transform
$\widehat{\psi}(\vec{ k})$ is the ellipse with  semi-axes $\Delta
k_\mathrm{x}$ and $\Delta k_\mathrm{y}$ and  center at the point
$\vec{ k} = (\overline{k_\mathrm{x}}, 0)$ (see Figure
\ref{pic-supp1}).

The pairs of  widths $\Delta x, \Delta k_\mathrm{x}$ and $\Delta y,
\Delta k_\mathrm{y}$  satisfy the Heisenberg uncertainty relation,
which holds for any function $\psi$:
\begin{equation}\label{Heisenberg}
 \Delta k_\mathrm{x} \,\Delta x \geq \frac{1}{2}, \,\,\,
\Delta k_\mathrm{y} \,\Delta y \geq \frac{1}{2}.
\end{equation}
Equality holds only for the Morlet wavelet (see, for example
\cite{Antoine-Book}), which is the asymptotics of the Gaussian Wave
Packet (\ref{packet3}) as $p \to \infty$.

Our purpose here is to study the dependence of the widths and  the
means of the wavelet $\psi$ and the uncertainty relation on the
parameters.   Taking $\varepsilon$ as the unit of  distance, we
rewrite the wavelet in terms of the dimensionless coordinates $x' =
x/\varepsilon$, $y' = y /\varepsilon ,$  the time $t' =
t/\varepsilon,$ and two parameters $p$ and $\varepsilon/\gamma$ :
\begin{equation}
\frac{\psi(\vec{r}')}{\sqrt{\varepsilon}} = \frac{(p s)^{\nu}
K_{\nu}(ps)}{\sqrt{x' + ct' -  i }} , \nonumber
\end{equation}
\begin{equation}
s = \left[1 -   i \frac{\varepsilon}{\gamma} \left(x' - ct' +
\frac{y'^2}{x' + ct' -  i } \right) \right]^{1/2}.
\label{params-repl}
\end{equation}
When $p$ is large,  the parameter $p$ and the ratio
$\varepsilon/\gamma$ can be interpreted with the help of the widths
of the wavelet Morlet (\ref{Morlet}), which were denoted by
$\sigma_\mathrm{x},\, \sigma_\mathrm{y},$ and the mean of the
longitudinal spatial frequency of its Fourier transform, which is
$\kappa.$ The ratio $\varepsilon/\gamma = \sigma_\mathrm{y} /
\sigma_\mathrm{x} $ characterizes the shape of the essential
numerical support of the wavelet, the product $\kappa
\sigma_\mathrm{x}  = \sqrt{p}$ is the number of  wavelengths  on the
width $\Delta x,$ the product $\kappa \sigma_\mathrm{y}  =
\sqrt{\kappa \varepsilon /2}$ is the number of  wavelengths on the
width $\Delta y$.

\begin{figure}[htbp]
\centering \subfloat[Widths of the Gaussian Wave Packet
$\psi(\vec{r})$ in relation to the widths of the Morlet wavelet,
solid lines are for $\Delta x / \sigma_\mathrm{x}$, dashed lines are
for $\Delta y / \sigma_\mathrm{y}$: line (1) is for $\varepsilon /
\gamma = 1/3$, line (2) is for $\varepsilon / \gamma = 2/3$, line
(3) is for $\varepsilon / \gamma = 2$.]{
\includegraphics[keepaspectratio,
width=0.7\textwidth]{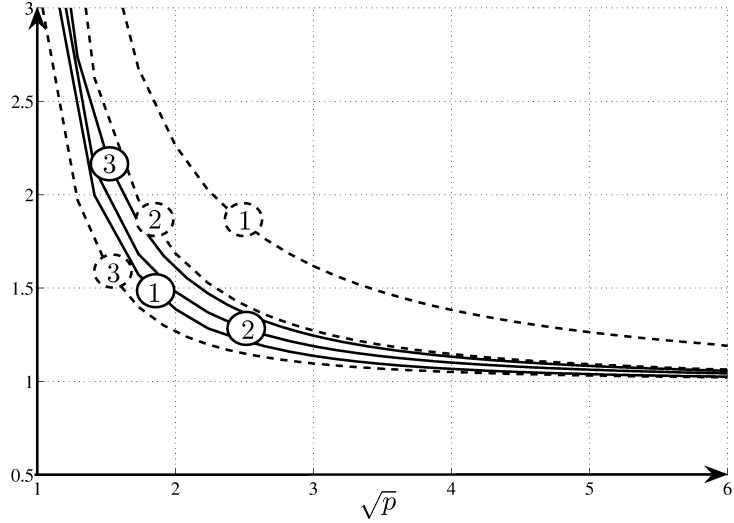} \label{width-vk:cord} }  \\
\subfloat[Widths of the Fourier transform of the Gaussian Wave
Packet $\widehat{\psi}(\vec{k})$ in relation to the widths of the
Fourier transform of the Morlet wavelet, solid lines are for $\Delta
k_\mathrm{x} / \sigma_{\mathrm{k}_\mathrm{x}}$, dashed lines are for
$\Delta k_\mathrm{y} / \sigma_{\mathrm{k}_\mathrm{y}}$: line (1) is
for $\varepsilon / \gamma = 1/3$, line (2) is for $\varepsilon /
\gamma = 2/3$, line (3) is for $\varepsilon / \gamma = 2$.]{
\includegraphics[keepaspectratio,
width=0.7\textwidth]{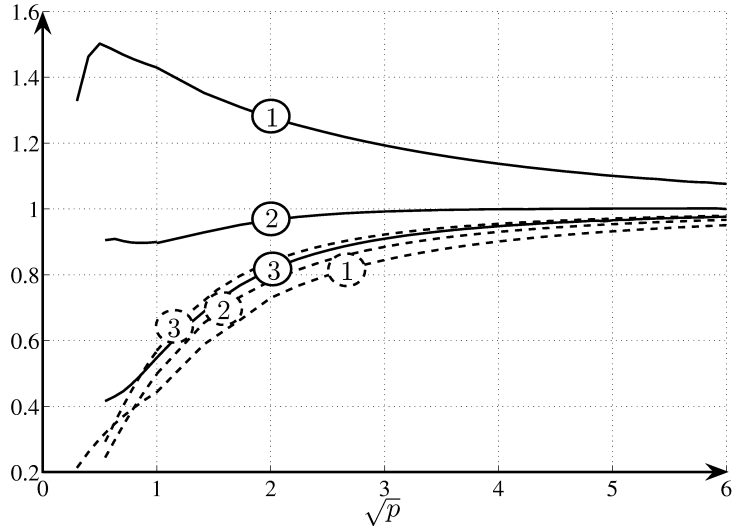} \label{width-vk:four}}
\caption{The ratios of the widths of the wavelet $\psi$ to the width
of the Morlet wavelet are plotted versus $\sqrt{p}$ in spatial (a)
and spatial frequency (b) domains for several values of $\varepsilon
/\gamma.$ } \label{width-vk}
\end{figure}

\begin{figure}[htbp]
\centering \subfloat[Widths of the Gaussian Wave Packet
$\psi(\vec{r})$ in relation to the widths of the Morlet wavelet,
solid lines for $\Delta x / \sigma_\mathrm{x}$, dashed lines are for
$\Delta y / \sigma_\mathrm{y}$: line (1) is for $2\kappa\varepsilon
= 4$, line (2) is for $2\kappa\varepsilon = 8$, line (3) is for
$2\kappa\varepsilon = 64$.]{
\includegraphics[keepaspectratio,
width=0.7\textwidth]{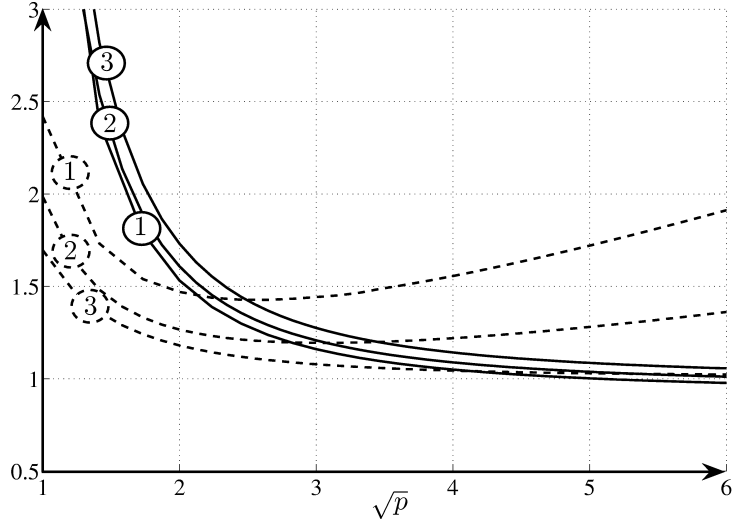} \label{width-vg:cord} } \\
\subfloat[Widths of the Fourier transform of the Gaussian Wave
Packet $\widehat{\psi}(\vec{k})$ in relation to the widths of the
Fourier transform of the Morlet wavelet, solid lines are for $\Delta
k_\mathrm{x} / \sigma_{\mathrm{k}_\mathrm{x}}$, dashed lines are for
$\Delta k_\mathrm{y} / \sigma_{\mathrm{k}_\mathrm{y}}$: line (1) is
for $2\kappa\varepsilon = 4$, line (2) is for $2\kappa\varepsilon =
8$, line (3) is for $2\kappa\varepsilon = 64$. ]{
\includegraphics[keepaspectratio,
width=0.7\textwidth]{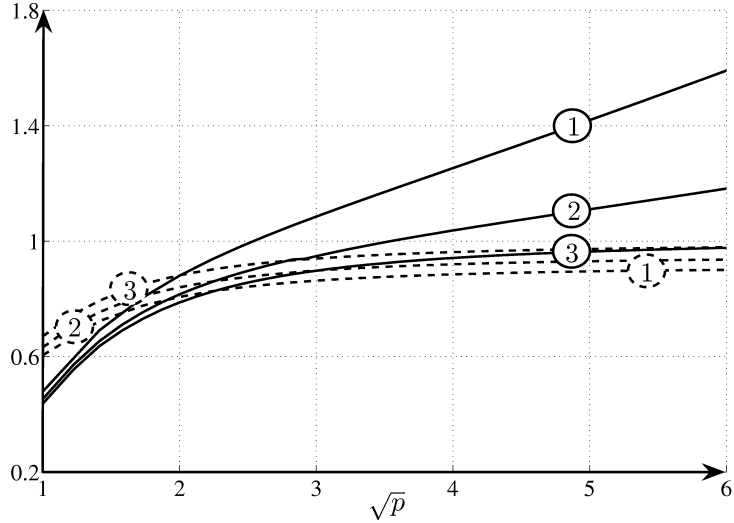} \label{width-vg:four}}
\caption{The relative widths of the wavelet $\psi$ are plotted
versus $\sqrt{p}$ in spatial (a) and spatial frequency (b) domains
for several values of $2\kappa\varepsilon $. } \label{width-vg}
\end{figure}

\begin{figure}[htbp]
\centering
\includegraphics[keepaspectratio,
width=0.7\textwidth]{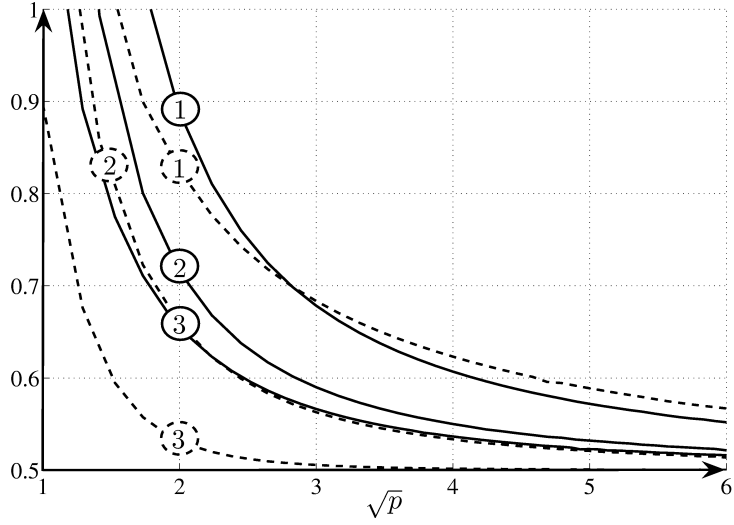}  \caption{The left-hand side of
the uncertainty relation for the Gaussian Wave Packet is plotted
versus $\sqrt{p}$, solid lines are for $ \Delta k_\mathrm{x}\Delta
x$, dashed lines are for $\Delta k_\mathrm{y} \Delta y$: line (1) is
for $\varepsilon / \gamma = 1/3$, line (2) is for $\varepsilon /
\gamma = 2/3$, line (3) is for $\varepsilon / \gamma = 2$.  }
\label{vk-uncert}
\end{figure}

\begin{figure}[htbp]
\centering
\includegraphics[keepaspectratio,
width=0.7\textwidth]{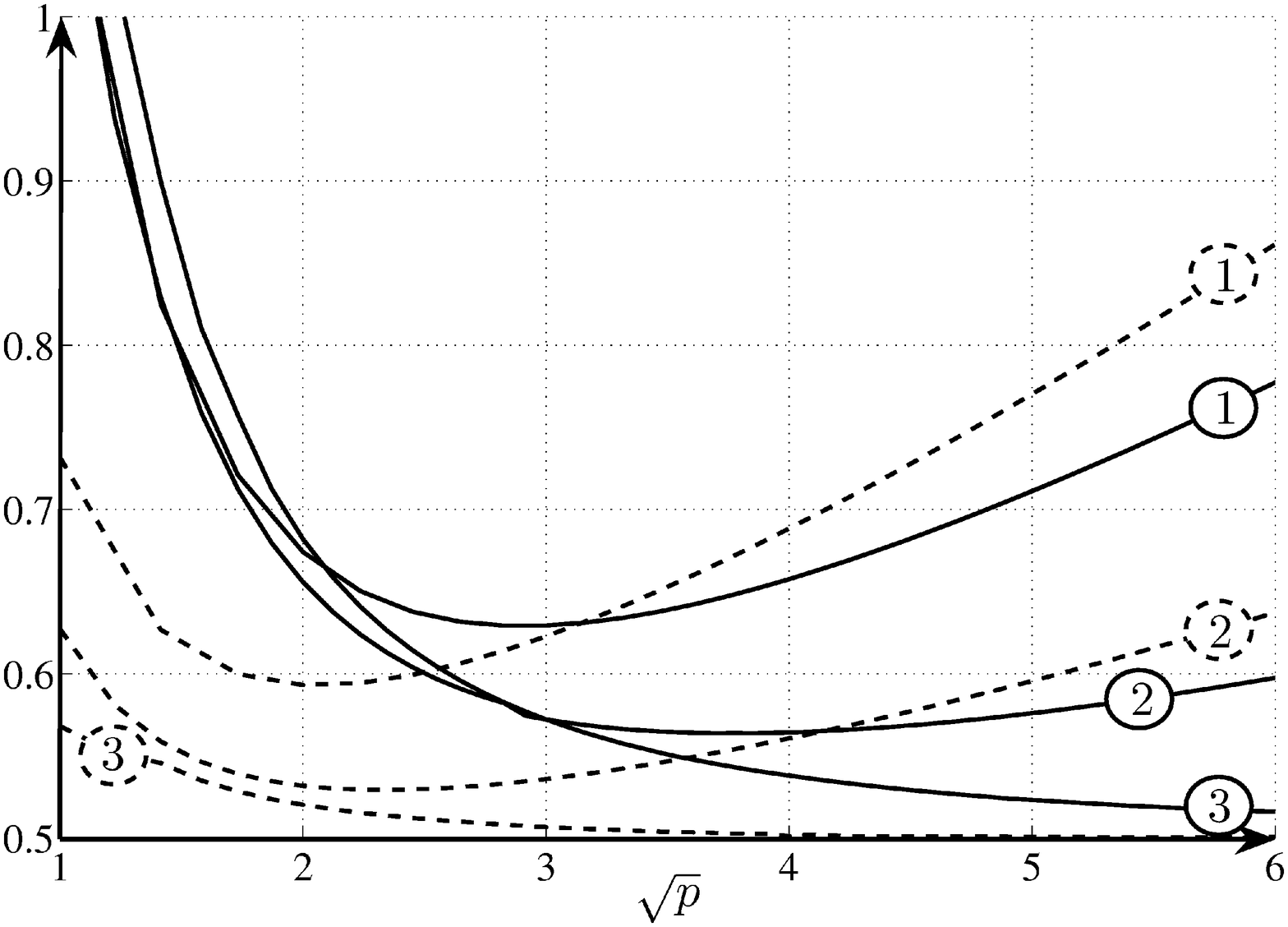}  \caption{The left-hand side of
the uncertainty relation for the Gaussian Wave Packet is plotted
versus $\sqrt{p}$, solid lines are for $ \Delta k_\mathrm{x} \Delta
x$, dashed lines are for $ \Delta k_\mathrm{y} \Delta y$: line (1)
is for $2\kappa\varepsilon = 4$, line (2) is for $2\kappa\varepsilon
= 8$, line (3) is for $2\kappa\varepsilon = 64$.  }
\label{vg-uncert}
\end{figure}

We have calculated numerically the widths of the  wavelet
(\ref{packet3}), its Fourier transforms, and the left-hand side of
the uncertainty relations (\ref{Heisenberg}) as functions of $p = 2
\kappa \gamma$ in two cases. In the first case, the parameter
$\kappa$ varies, the parameters $\varepsilon$ and $\gamma$ are kept
constant. When $p$ is large, it can be interpreted  as follows: the
shape of the essential numerical support of the Morlet wavelet is
kept constant and the number of  wavelengths along the transverse
and longitudinal widths increases with $\kappa.$ The results are
plotted in Figures \ref{width-vk}, \ref{vk-uncert}. In the second
case, $\gamma$ varies, the parameters $\varepsilon$ and $\kappa$ are
kept constant. For large $p,$ this means that  the number of
wavelengths within the transverse width is kept fixed but the
longitudinal width increases with increase of $\gamma$. The results
are plotted in Figures \ref{width-vg}, \ref{vg-uncert}. In all
cases, the number of  wavelengths on the width $\Delta x,$ i.e.,
$\sqrt{p}$ is laid off along the horizontal axis. It is the
parameter $\sqrt{p}$ that controls the asymptotic behavior of the
new wavelet, because the correcting term in (\ref{Morlet}) is of
 order  $ O  (1/\sqrt{p})$ when $\alpha= {1 \over 2}$. In view of
comparing the widths with their asymptotics, the relatives widths,
i.e., $\Delta_\mathrm{x} / \sigma_\mathrm{x},$ and so on, are
plotted as ordinates in Figures \ref{width-vk}, \ref{width-vg}.

Figure \ref{width-vk} shows that all the widths tend to their
asymptotic values as $\sqrt{p} \to \infty,$ the parameter
$\varepsilon/\gamma$ being fixed.  The widths of the wavelet in the
coordinate domain are larger than their asymptotics
$\sigma_\mathrm{x},$ $\sigma_\mathrm{y}$ (Figure
\ref{width-vk:cord}). The widths of the wavelet in the spatial
frequency domain $\Delta k_\mathrm{x},$ $\Delta k_\mathrm{y}$ may be
both smaller or larger than the same widths of the Morlet wavelet
$\sigma_{\mathrm{k}_\mathrm{x}} = 1/(2 \sigma_\mathrm{x}),$ \,
$\sigma_{\mathrm{k}_\mathrm{y}} = 1/(2 \sigma_\mathrm{y})$ (Figure
\ref{width-vk:four}). The larger  the parameter
$\varepsilon/\gamma,$ the closer to the saturation  the Heisenberg
uncertainty relation (see Figure \ref{vk-uncert}) and the smaller
$\Delta \mathrm{y} / \sigma_\mathrm{y}$ (Figure \ref{width-vk:cord})
and $\Delta \mathrm{k}_\mathrm{x} / \sigma_{\mathrm{k}_\mathrm{x}}$
( Figure \ref{width-vk:four}). However $\Delta \mathrm{x} /
\sigma_\mathrm{x}$ and $\Delta \mathrm{k}_\mathrm{y} /
\sigma_{\mathrm{k}_\mathrm{y}}$ increase somewhat with  increase in
$\varepsilon/\gamma,$ remaining smaller than their counterparts.

To interpret  Figures \ref{width-vg}, \ref{vg-uncert}, we note that
the rate of  convergence of the wavelet to the Morlet asymptotics is
determined by terms of order  $p^{-\alpha} \gamma / \varepsilon =
(2\kappa)^{-\alpha} \gamma^{1-\alpha}/ \varepsilon,$ which must be
small. However these terms  increase with  $\gamma$ if $\varepsilon$
and $\kappa$ are fixed.

\subsection{Directional properties of the wavelet $\psi$}

A wavelet is called directional (see \cite{Antoine-Book} for greater
detail) if the  essential numerical support of its Fourier transform
lies in a convex cone in $\vec{k}$ space with its vertex at the
origin and the angle $\alpha$ at the vertex  (See Figure
\ref{pic-supp1}). Using the asymptotic formula (\ref{Morlet}), we
can easily prove that the Gaussian Wave Packet is a directional
wavelet when $p$ is large enough. This follows from the fact that
$\overline{k_\mathrm{x}} = \kappa = p/(2\gamma)$, $\Delta
k_\mathrm{x} = \sqrt{p}/(4\gamma) $ as $p \to \infty$, $\gamma$ and
$\varepsilon$ are fixed. Thus the inequality
$\overline{k_\mathrm{x}}
> \Delta k_\mathrm{x}$ holds, which ensures that the origin lies outside the
ellipse.

\begin{figure}[htbp]
\centering
\includegraphics[keepaspectratio,
width=0.7\textwidth]{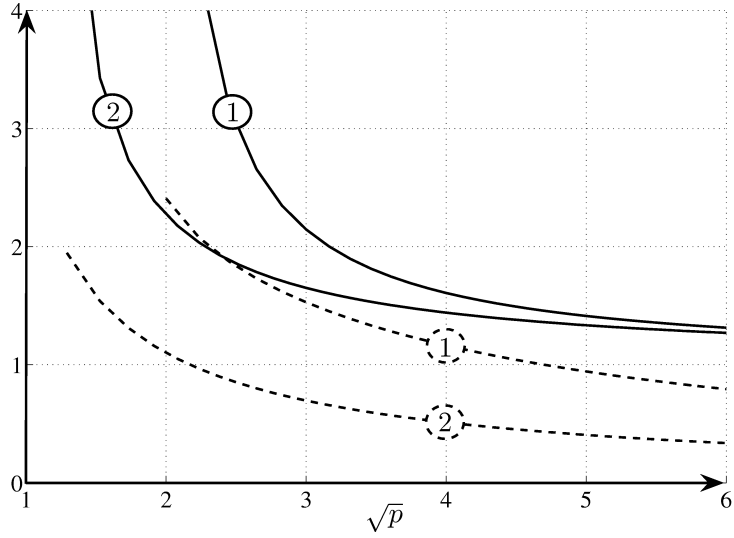}  \caption{Scale and angle
resolving powers for the Gaussian Wave Packet are plotted versus
$\sqrt{p}$, solid lines are for the SRP, dashed lines are for ARP:
line (1) is for $\varepsilon / \gamma = 1/3$, line (2) is for
$\varepsilon / \gamma = 2$.} \label{vk-aprspr}
\end{figure}

\begin{figure}[htbp]
\centering
\includegraphics[keepaspectratio,
width=0.7\textwidth]{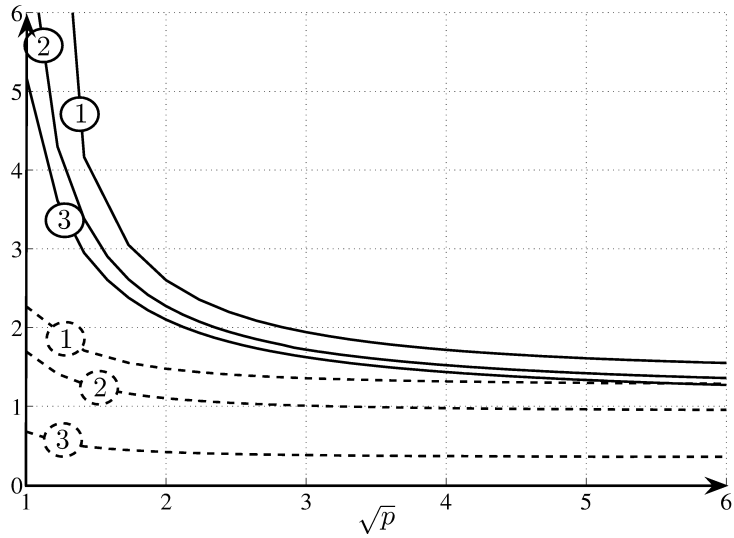}  \caption{Scale and angle
resolving powers for the Gaussian Wave Packet are plotted versus
$\sqrt{p}$, solid lines are for the SRP, dashed lines are for ARP:
line (1) is for $2\kappa\varepsilon = 4$, line (2) is for
$2\kappa\varepsilon = 8$, line (3) is for $2\kappa\varepsilon =
64$.} \label{vg-aprspr}
\end{figure}

We can calculate numerically the scale resolving power (SRP) and the
angular resolving power (ARP)  if the wavelet is directional. These
quantities  are especially important for  numerical calculations,
mainly in determining the minimal sampling grid for a lossless
reconstruction of the image (for more detail, see
\cite{Antoine-article}, \cite{Antoine-Book}). The resolving powers
are determined by the expressions
\begin{equation}\label{SRP}
SPR(\psi) = \frac{\overline{k_\mathrm{x}} + \Delta
k_\mathrm{x}}{\overline{k_\mathrm{x}} - \Delta k_\mathrm{x}}
\end{equation}
\begin{equation}\label{ARP}
APR(\psi) = 2 \, \mathrm{arccot}
\frac{\sqrt{\left(\overline{k_\mathrm{x}}\right)^2 - \left(\Delta
k_\mathrm{x}\right)^2}}{\Delta {k_\mathrm{y}}} = \alpha.
\end{equation}
As $SRP$ tends to $1$ and $ARP$ tends to $0$, the wavelet becomes
more sensitive to small singularities and to angular details of the
analyzed data \cite{Antoine-Book}. The  dependance of the angle and
scale resolving powers was calculated by using MATLAB and is
presented in Figure {\ref{vk-aprspr}} and Figure \ref{vg-aprspr} for
the cases of varying $\kappa$ and $\gamma,$ respectively.

\section{Multidimensional case \label{sec-manydim}}

\subsection{Definition and main properties}

Here we give a generalization of the wavelets constructed in
previous sections  to the case of many spatial dimensions. The
counterpart of (\ref{packet3}) in $\mathbb{R}^n$ is
\begin{equation}
\psi(\vec{r}) = \frac{ \sqrt{2/\pi} \qquad (ps)^{\nu} K_{\nu}(ps) }
{ (x_1 + c t -  i \varepsilon_2)^{1/2}(x_1 + c t -  i
\varepsilon_3)^{1/2}\cdot \ldots \cdot(x_1 + c t -  i
\varepsilon_{\mathrm{n}})^{1/2}}  \label{packetn3}
\end{equation}
where  $\vec{ r}=(x_1,x_2,x_3,\ldots,x_{\mathrm{n}}),$ $p$,
$\gamma,$ $\varepsilon_2, \ldots, \varepsilon_\mathrm{n}$ are
positive parameters, the function $s$ is defined by (\ref{def-s}),
and it depends on $\vec{r}, t,$ and on the parameters via $\theta,$
which is as follows:
\begin{equation}
\theta = x_1 - ct + \frac{x_2^2 }{x_1 + c t -  i \varepsilon_2} +
\frac{x_3^2 }{x_1 + c t -  i \varepsilon_3} + \ldots +
\frac{x_{\mathrm{n}}^2 }{x_1 + c t -  i \varepsilon_\mathrm{n}}.
\label{thetan}
\end{equation}
If $t$ is  time, the function $\psi(\vec{ r})$ satisfies the wave
equation
\begin{equation}
\psi_{tt} - c^2 ( \psi_{x_1 x_1} + \psi_{x_2 x_2} + \ldots +
\psi_{x_\mathrm{n} x_\mathrm{n}} ) = 0, \qquad c = \mathrm{const}.
\label{wave-n}
\end{equation}
 If $\nu={1 \over 2},$ formula (\ref{packetn3}) is transformed to an exact
solution of (\ref{wave-n}), found in \cite{Kis-Per-many}:
\begin{equation}
\psi(\vec{r}) = \frac{\exp{\left( - p \sqrt{1 - { i
\theta}/{\gamma}} \right)}} {(x_1 + c t -  i  \varepsilon_2)^{1/2}
(x_1 + c t -  i  \varepsilon_3)^{1/2}\cdot \ldots \cdot(x_1 + c t -
 i \varepsilon_{\mathrm{n}})^{1/2} }. \label{packetn1}
\end{equation}
The exact solution (\ref{packetn3}) was mentioned in
\cite{Perel-Fialkovsky}. The Fourier transform of the wavelet
(\ref{packetn3}) is found in the Appendix. It reads
\begin{equation}
\widehat{\psi} (\vec{k}) = a_n \frac{p^{2\nu}}{\gamma^\nu} \left[
k ( k + k_1 )^{\nu+(n-1)/2} \right]^{-1} \times \nonumber \\
\end{equation}
\begin{equation}
\times \exp\left[ -  \frac{k + k_1}{2}\gamma - \frac{k_2^2
\varepsilon_2 + k_3^2 \varepsilon_3 + \ldots + k_{\mathrm{n}}^2
\varepsilon_{\mathrm{n}}} {2( k + k_1 )} - \frac{2 \kappa^2
\gamma}{k + k_1} -   i  k c t \right] , \label{Four-n}
\end{equation}
where $\vec{ k} = (k_1, k_2, \ldots, k_\mathrm{n})$ is a wave
vector, $k = |\vec{ k}|,$
\begin{equation}
a_n = (2\pi)^{n/2}   e ^{  i \pi(n-1)/4}.  \label{a-n}
\end{equation}

The axially symmetric wavelet follows from (\ref{packetn3}),
(\ref{thetan}) if $\varepsilon_2 = \varepsilon_3 = \ldots =
\varepsilon_\mathrm{n} \equiv \varepsilon.$ It reads
\begin{equation}
\psi(\vec{r}) = \sqrt{\frac{2}{\pi}} \, \frac{ (ps)^{\nu}
K_{\nu}(ps) }{ (x_1 + c t -  i \varepsilon)^{(n-1)/2 }}.
\label{packetn3-sym}
\end{equation}
Its Fourier transform is
\begin{equation}
\widehat{\psi}(\vec{k}) = a_n \frac{p^{2\nu}}{\gamma^\nu} \left[
k(k+k_1)^{\nu+(n-1)/2} \right]^{-1} \times \nonumber
\end{equation}
\begin{equation}
\times \exp\left[ - (k + k_1 )\frac{\gamma}{2} - (k - k_1
)\frac{\varepsilon}{2} - \frac{2\kappa^2\gamma}{k+k_1} -   i  k c t
\right] . \label{Four-n-Axsym}
\end{equation}
All the properties of (\ref{packet3}) are extended to
(\ref{packetn3}) with minor corrections, i.e., it is a smooth
function of  coordinates, its moments of any order are zero. The
same is true for its derivatives of any order with respect to
coordinates and  $t$. The modulus of (\ref{packetn3}) has a maximum
at the point $x_1=ct,$ $x_j=0, j=2,\ldots,n.$ It has asymptotics
(\ref{Morlet2})  as $p \to \infty,$ where
\begin{equation}
\psi_{\mathrm{beam}}(\kappa, \vec{r}) = \frac{ \exp{(  i  \, \kappa
\, \theta)} }{(x_1 + c t -  i \varepsilon_2)^{1/2} \ldots (x_1 + c t
-  i \varepsilon_{\mathrm{n}})^{1/2}}, \label{beamn}
\end{equation}
which is uniform in the domain
\begin{equation}
\frac{x_1-ct}{\gamma} =  O  \left(\frac{1}{p^{\alpha}}\right),
\qquad \frac{x_j}{ \sqrt{\gamma\varepsilon_j} } =  O
\left(\frac{1}{p^{\alpha}}\right), \qquad j = 2, 3, ..., n,
\label{estim-vicin-n}
\end{equation}
and ${1 \over 3} \leq \alpha \leq {1 \over 2}$. If the parameters $2
c t/\varepsilon_j,$ $p^{-\alpha}\gamma / \varepsilon_j ,\,j=2,..,n,$
are small, the asymptotics of the formula (\ref{Morlet2}) with
$\psi_{\mathrm{beam}}$ from (\ref{beamn}) represents  the Morlet
wavelet
\begin{equation}
\psi(\vec{r}) \sim \exp \left[  i \kappa (x_1 - c t)  - \frac{ (x_1
- c t)^2}{2\sigma_1^2} - \frac{x_2^2}{2\sigma_2^2} \ldots -
\frac{x_\mathrm{n}^2}{2\sigma_\mathrm{n}^2} \right] \times \nonumber
\end{equation}
\begin{equation}
\times \frac{C}{(-  i \varepsilon_2)^{1/2}(-  i \varepsilon_3)^{1/2}
\ldots (- i \varepsilon_\mathrm{n})^{1/2}}, \label{Morlet-nd}
\end{equation}
where $ \sigma_1^2 = { 4\gamma^2 }/{p },$ $ \sigma_2^2 = \gamma
\varepsilon_2/p, ...$ $ \sigma_\mathrm{n}^2 = \gamma
\varepsilon_\mathrm{n}/p $.

\subsection{Coefficient $C_{\psi}$}

For the usage of the Gaussian Wave Packet as a mother wavelet, we
need to calculate the coefficient $C_{\psi}$ (for example, see
\cite{Daubechies}, \cite{Antoine-Book}) defined by the formula
\begin{equation}\label{C_psi_N}
C_{\psi} = \frac{1}{(2\pi)^n} \int\limits_{\mathbb{R}^n}   d ^n
\vec{k} \frac{|\widehat{\psi}(\vec{ k})|^2}{|\vec{ k}|^n}.
\end{equation}
Although we cannot calculate this integral analytically in the
general case, we can simplify this expression. It should be noticed
that
\begin{eqnarray}
-\frac{(  i  c)^{n+1}}{n!} \int\limits_{-  i \infty}^{0}  d  t \,
t^n \, \exp(-  i  kct) = \frac{1}{k^{n+1}}. \nonumber
\end{eqnarray}
This equation makes it possible to treat the integrand in
(\ref{C_psi_N}) in the way
\begin{equation}\label{coef2}
\frac{|\widehat{\psi}(\vec{k})|^2}{|\vec{k}|^n} = - \frac{  e ^{-
 i  \pi(n-1)/4} } {(2\pi)^{n/2}} \frac{\gamma^{(n-1)/2}}{p^{n-1}}
\frac{(  i  c)^{n+1}}{n!} \int\limits_{-  i \infty}^{0}   d  t \,
t^n \, \widehat{\psi}_2(\vec{k}, t).
\end{equation}
Here we denote  the Gaussian Wave Packet with parameters $(\gamma,
\varepsilon_2, ... \varepsilon_\mathrm{n}, \kappa, \nu)$ by $\psi$
and  the packet with  parameters $(2\gamma, 2\varepsilon_2, ...,
2\varepsilon_\mathrm{n}, \kappa, 2\nu+(n-1)/2)$  by $\psi_2$. Then
we can view the integral (\ref{C_psi_N}) as a Fourier inverse
transform calculated at the point $\vec{ r}=0$. Substituting the
expression on the right-hand side of (\ref{coef2}) into
(\ref{C_psi_N}) and  changing the order of  integrals, we get
\begin{equation}\label{C_Psi_N_Result}
C_{\psi} = - \frac{  e ^{-  i  \pi(n-1)/4}} {(2\pi)^{n/2}}
\frac{\gamma^{(n-1)/2}}{p^{n-1}} \frac{(  i  c)^{n+1}}{n!}
\int\limits_{-  i \infty}^{0}   d  t \, t^n \, \psi_2(0, t).
\end{equation}

In the case where $n=3$ and $\varepsilon = \gamma,$ we can
explicitly calculate the coefficient $C_{\psi}$ from the formula
(\ref{C_psi_N}) if the coefficient $2\nu$ is a nonnegative integer.
We calculate this integral in the spherical system of coordinates
and use the integral representation of McDonald's function
$K_{\lambda}(z)$ (see \cite{Abramovitz-Stigan}). We denote $k =
|\vec{k}|$. As a result, we obtain
\begin{equation}
C_{\psi} = \int\limits_{\mathbb{R}^3}   d ^3\vec{k} \,
\frac{|\widehat{\psi}(\vec{k})|^2}{|\vec{k}|^3} \nonumber
\end{equation}
\begin{equation}
= (2\pi)^3 \frac{p^{4\nu}}{\gamma^{2\nu}} \int\limits_{\mathbb{R}^3}
d ^3\vec{k} \,\, \frac{1}{|\vec{k}|^5(|\vec{k}| +
k_\mathrm{x})^{2\nu + 2}} \exp\left[ -2\gamma|\vec{k}| -
\frac{4\kappa^2 \gamma}{|\vec{k}| + k_\mathrm{x}} \right] \nonumber
\end{equation}
\begin{equation}
= (2\pi)^4 \frac{p^{4\nu}}{\gamma^{4\nu}} \sum\limits_{m=0}^{2\nu}
\, \frac{(2\nu)!}{m!} \, 2^{-4\nu + m -1} \, \frac{\kappa^{-4\nu + m
- 5}}{\gamma^{-m+1}} \, K_{m+3}(4\kappa\gamma). \label{C_psi_3d}
\end{equation}

\section{New wavelets and pulse complex sources \label{sec-pulssour}}

In  previous sections, we discussed  properties of a Gaussian packet
viewed at a fixed moment of time as a wavelet.  Now we are going to
review  properties of a Gaussian packet as a physical wavelet. The
term "physical wavelet" has been first suggested by G. Kaiser in
\cite{Kaiser}. These wavelets have been obtained as a sum of fields
of two point sources with a special time dependence.

First we consider two wave equations with  moving $\delta$-sources
on the right-hand sides:
\begin{equation}
u^{\pm}_{tt} - c^2 (u^{\pm}_{xx} + u^{\pm}_{yy}) =  \pm\phi(q,t)
\delta ( x + c t ) \, \delta(y), \nonumber
\end{equation}
\begin{equation}
\phi(q, t) = A \sqrt{q} \exp(-2  i  q c t), \qquad A = -  e ^{- i
\pi/4} 4 \sqrt{\pi} c^2. \label{eq_pair}
\end{equation}
The equation, say, for  $u^{+}$ can be solved if we seek a solution
in the form
\begin{equation}
u^{+}(x,y,t) = \exp(  i  q\alpha) g(\beta, y), \qquad \alpha = x -
ct, \qquad \beta = x + ct.
\end{equation}
This leads to the Schr\"{o}dinger equation for the function
$g(\beta, y)$:
\begin{equation}
4  i  q g_{\beta} + g_{yy} =  -4\sqrt{\pi}   e ^{- i \pi/4} \sqrt{q}
\delta(\beta) \delta(y).
\end{equation}
The solution of this equation is merely a well-known fundamental
solution for the Schr\"{o}dinger equation. So we get
\begin{equation}\label{sol_1}
u^{+}(\vec{r}, t) = \frac{\Theta(x + ct)}{\sqrt{x + ct}} \exp\left[
 i  q\left(x - ct + \frac{y^2}{x + ct} \right) \right],
\end{equation}
where $\Theta$ is the Heaviside step function. We interpret it as a
field of a source moving in the negative direction of the $x$ - axis
at a speed $c,$  emitting an elementary pulse $\phi(q,t)$. The
solution does vanish in front of the moving source, i.e., for
$x<-ct,$ this is why it is called retarded. We denote  the advanced
solution, which does  vanish behind  the moving point, by $u^{-}$.
It reads
\begin{equation}\label{sol_11}
u^{-}(\vec{r}, t) = \frac{\Theta(- x - ct)}{\sqrt{x + ct}}
\exp\left[  i  q\left(x - ct + \frac{y^2}{x + ct} \right) \right].
\end{equation}
It can be interpreted as a field absorbed by the source moving in
the negative direction of the $x$ - axis at a speed $c$ and
producing  the current $-\phi(t)$ in this source.

We note that $u = u^{+} + u^{-}$ is a sourceless solution, i.e., it
satisfies the homogeneous wave equation, because $\Theta(x + ct) +
\Theta(-x -ct) = 1$.  However, it has a singularity at the point $x
= - ct$. In order to eliminate it, we shift $x$ by $ -  i
\varepsilon/2, \,\, \varepsilon
> 0$ and $t$ by $-  i \varepsilon/(2c)$ to the
complex plane. This leads to a  nonstationary Gaussian beam
(\ref{beam}):
\begin{equation}
u(\vec{r}, t) = \psi_{\mathrm{beam}}(q, \vec{r}) = \frac{1}{\sqrt{x
+ ct -  i \varepsilon}} \exp\left[  i  q\left(x - ct + \frac{y^2}{x
+ ct -  i  \varepsilon} \right) \right].
\end{equation}

We get a Gaussian beam, using the sum of  fields of  moving sources,
one of which emits and the other absorbs a pulse $\phi(q,t)$ that is
proportional to $\exp(-2 i  ctq).$  This result makes it possible to
use an arbitrary function of  time $t$ instead of an exponent on the
right-hand side of (\ref{eq_pair}), decomposing it into a Fourier
integral. We apply
\begin{equation}
\Phi(t) = \int\limits_{0}^{\infty}  d  q \, \mathcal{F}(q) \, \phi(
q, t) \equiv B \, \sigma^{\nu-1/2} K_{\nu-1/2}(\sigma),
\label{impulse-source}
\end{equation}
\begin{equation}
\sigma = 2\kappa\gamma (1 + 2  i  c t/\gamma)^{1/2} \label{sig},
\qquad B = -4 c^2  e ^{-  i \pi/4} \frac{p}{\sqrt{\gamma}},
\end{equation}
\begin{equation}
\mathcal{F}(q) =  a \, q^{-\nu-1} \, \exp\left[ -\gamma \left( q +
\frac{\kappa^2}{q} \right)\right]
\end{equation}
instead of $\phi(q,t)$ as input source functions  in (\ref{eq_pair})
and obtain (\ref{packet3}) instead of $\psi_{\mathrm{beam}}(q, \vec{
r})$.

\section{ Conclusions}

We suggest that we should consider an exponentially localized
solution of the wave equation in several spatial dimensions, found
earlier, from the point of view of  continuous wavelet analysis. We
investigate its properties as a mother wavelet and as a solution of
the wave equation in view of its further application to  the study
of local properties and singularities of  acoustic or optic fields.

We show that, depending on the parameters, the solution represents a
short pulse of one oscillation, or  a wave packet with the Gaussian
envelop filled with oscillations, or  a nonstationary Gaussian beam
multiplied by a cutoff function.

The solution for a fixed time is a multidimensional  wavelet with
all zero moments. Its Fourier transform is calculated explicitly,
and it is exponentially localized. The widths of the new wavelet in
the position domain and in the spatial frequency domain, and the
Heisenberg uncertainty relation are numerically investigated.

\section*{Acknowledgments}
M.Sidorenko was supported by The
Dmitry Zimin 'DYNASTY' Foundation. \\

\appendix
\section{ Fourier transform of the Gaussian Wave Packet}

\subsection{Fourier transform of the two-dimensional packet}

The Fourier transform of the wavelet (\ref{packet3}) can be
calculated using the Fourier decomposition of the Gaussian Wave
Packet (\ref{packet3}) in terms of the Gaussian beams (\ref{beam})
with the help of the formula
\begin{equation}
\psi(\vec{r}) = \int\limits_{0}^{\infty}  d  q \, \mathcal{F}(q) \,
\psi_{\mathrm{beam}}(q, \vec{r}). \label{decomp-F}
\end{equation}
To obtain this decomposition,  we use the formula (see
\cite{Abramovitz-Stigan})
\begin{equation}
\sqrt{\frac{2}{\pi}}\, (ps)^{\nu} \,K_{\nu}(ps) =
\int\limits_{0}^{\infty}  d  q \, \mathcal{F}(q) \, \exp( \,  i  \,
q \, \theta), \label{decomp-F2} \\
\end{equation}
\begin{equation}
\mathcal{F}(q) =  a \, q^{-\nu-1} \, \exp\left[ -\gamma \left( q +
\frac{\kappa^2}{q} \right)\right], \label{calF} \\
\end{equation}
\begin{equation}
a = \frac{1}{\sqrt{2\pi}} \frac{p^{2\nu}}{(2\gamma)^\nu}
,\label{def-F}
\end{equation}
where the left-hand side (\ref{decomp-F2}) depends on $\theta$ via
$s$ from (\ref{def-s}). We note that this relation is valid in the
upper complex half-plane $\mathrm{Im}(\theta) > 0,$  where $\theta$
lies when $x,y$ and time $t$ are real.  Dividing both sides of
(\ref{decomp-F2}) by the term $\sqrt{x + ct -  i \varepsilon},$ we
obtain (\ref{decomp-F}). So the Fourier transform of the Gaussian
Wave Packet is determined by the formula
\begin{equation}
\widehat{\psi}(\vec{k}) = \int\limits_{0}^{\infty}  d  q \,
\mathcal{F}(q) \, \widehat{\psi}_{\mathrm{beam}}(\vec{k}, q).
\label{decomp-Four-beam}
\end{equation}
The Fourier transform of the Gaussian beam (\ref{beam}) is found to
be
\begin{equation}
\widehat{\psi}_{\mathrm{beam}}(\vec{k})=\int\limits_{\mathbb{R}^2}
\,  d ^2\vec{r} \, \frac{\exp\left(-  i \vec{k} \cdot \vec{r}
\right)} {\sqrt{x + ct -  i \varepsilon}} \exp\left[  i  q \left(x -
ct +
\frac{y^2}{x + ct -  i \varepsilon} \right) \right]  \nonumber \\
\end{equation}
\begin{equation}
= \int\limits_{-\infty}^{\infty}  d  x \, \exp( - i  k_\mathrm{x} x)
\, \,\exp[ i  q (x-ct)] \, I_\mathrm{y}(q,x), \label{Iy-Ix}
\end{equation}
where
\begin{equation}
I_\mathrm{y}(q, x) = \int\limits_{-\infty}^{+\infty}   d  y \,
\frac{1}{\sqrt{x + ct -  i \varepsilon}} \exp\left(-  i k_\mathrm{y}
y +  i  q \frac{y^2}{x + ct -  i \varepsilon}\right) \, \nonumber
\end{equation}
\begin{equation}
= \sqrt{\frac{\pi}{q}} \,  e ^{ i \pi/4 } \, \exp\left[ - i
\frac{k_\mathrm{y}^2}{4q} (x + ct -  i  \varepsilon) \right].
\label{Ix}
\end{equation}
Inserting (\ref{Ix}) into (\ref{Iy-Ix}), we obtain
\begin{equation}
\widehat{\psi}_{\mathrm{beam}}(\vec{k}) =
\int\limits_{-\infty}^{\infty}  d  x \, \exp( - i  k_\mathrm{x} x )
\, \exp\left[  i  q( x - ct ) -  i \frac{k_\mathrm{y}^2}{4q} (x + ct
-  i \varepsilon) \right]  e ^{ i \pi/4} \sqrt{\frac{\pi}{q}}
\nonumber \\
\end{equation}
\begin{equation}
= 2\pi \, \exp{\left[- i  \left( q + \frac{k_\mathrm{y}^2}{4q}
\right) ct - \frac{k_\mathrm{y}^2 \varepsilon}{4q}  \right] } \,
\delta\left(-k_\mathrm{x} + q -\frac{k_\mathrm{y}^2}{4q} \right) e
^{ i \pi/4} \sqrt{\frac{\pi}{q}} \label{Four-beam-interm}.
\end{equation}
In view of  further calculations of the integral with respect to
$q,$ we modify the $\delta$ - function in (\ref{Four-beam-interm}).
The roots of the argument of the $\delta$ - function, i.e., the
roots of the equation
\begin{equation}
-k_\mathrm{x} + q -\frac{k_\mathrm{y}^2}{4q} = 0
\end{equation}
are $q_{1,2} = (k_\mathrm{x} \pm k)/2$, $k = |\vec{ k}|.$ The
negative root does not contribute to the integral, because  ${\cal
F}(q, \varepsilon) \equiv 0$ if $ q < 0.$ Taking into account the
relations
\begin{equation}
\delta\left(-k_\mathrm{x} + q -\frac{k_\mathrm{y}^2}{4q} \right) =
\frac{4 q^2}{k_\mathrm{y}^2 + 4 q^2} \, \delta\left( q -
\frac{k_\mathrm{x} + k}{2} \right),
\end{equation}
\begin{equation}
\frac{k_\mathrm{y}^2}{4q} + q = \frac{k^2 - k_\mathrm{x}^2}{2 (k +
k_\mathrm{x}) } + \frac{k + k_\mathrm{x}}{2 } = k, \qquad \frac{4
q^2}{k_\mathrm{y}^2 + 4 q^2} = \frac{q}{k} = \frac{k_\mathrm{x} +
k}{2 k},
\end{equation}
we obtain
\begin{equation}
\widehat{\psi}_{\mathrm{beam}}(\vec{ k}) = B(\vec{ k}) \,
\delta\left( q - \frac{k_\mathrm{x} + k}{2} \right),
\end{equation}
\begin{equation}
B(\vec{k}) = 2\pi \, \exp{\left(- i  k c t - \frac{k_\mathrm{y}^2
\varepsilon}{2 (k + k_\mathrm{x}) } \right)} \, \frac{k +
k_\mathrm{x}}{2 k}  e ^{ i \pi/4 } \sqrt{\frac{2 \pi}{k +
k_\mathrm{x}}}.\label{F-b-2}
\end{equation}
After inserting (\ref{F-b-2}) into (\ref{decomp-Four-beam}), the
integral disappears and the resulting expression contains
$\mathcal{F}[(k + k_\mathrm{x})/2],$ which is as follows:
\begin{equation}
\mathcal{F} \left(\frac{k + k_\mathrm{x}}{2}\right) = a \;
\frac{2^{\nu+1}}{(k + k_\mathrm{x})^{\nu+1} } \, \exp{\left(- \frac{
k + k_\mathrm{x} }{2} \, \gamma \, - \, \frac{2 \kappa^2 \gamma}{ k
+ k_\mathrm{x}} \right)}.
\end{equation}
Taking into account
\begin{equation}
\frac{k_\mathrm{y}^2 \varepsilon}{2(k+k_\mathrm{x})} = \frac{k -
k_\mathrm{x}}{2} \varepsilon,
\end{equation}
we obtain
\begin{equation}
\widehat{\psi}(\vec{k}) =  \mathcal{F} \left(\frac{k +
k_\mathrm{x}}{2}\right)\, B(\vec{k}), \nonumber
\end{equation}
\begin{equation}
\widehat{\psi}(\vec{k})  = 2\pi \,  e ^{ i \pi/4}
\frac{p^{2\nu}}{\gamma^\nu} \, \left[k (k + k_x)^{\nu+1/2}
\right]^{-1}  \nonumber
\end{equation}
\begin{equation}
\times \exp\left[- i  k c t - (k - k_\mathrm{x})
\frac{\varepsilon}{2} - (k + k_\mathrm{x})\frac{\gamma}{2} - \frac{2
\kappa^2 \gamma}{ k + k_\mathrm{x}} \right]. \label{gen-Four-2}
\end{equation}

\subsection{Fourier transform of a multidimensional Gaussian Wave Packet}

In the multidimensional case, the Fourier transform of a Gaussian
beam must be re-calculated. We  put $\varepsilon_2 \, = \,
\varepsilon_3 \, = \, ... \, = \,\varepsilon_\mathrm{n} \, = \,
\varepsilon$ from (\ref{packetn3}) for simplicity. Instead of
$I_\mathrm{y},$  the formula (\ref{Iy-Ix}) will contain  the product
$I_{\mathrm{x}_2}\cdot I_{\mathrm{x}_3} \ldots
I_{\mathrm{x}_\mathrm{n}},$ where $n-1$ is the number of coordinates
that are transverse to the direction of propagation. We obtain a
formula for $I_{\mathrm{x_k}},$ $k=2,...,n,$ by replacing $y$ by
$x_\mathrm{k}$ and $x$ by $x_1$ in (\ref{Ix}). The analog of
(\ref{Four-beam-interm}) will contain the sum $k_2^2 + k_3^2 +
\ldots + k_\mathrm{n}^2 = k^2 - k_1^2$ instead of the term
$k_\mathrm{y}^2 = k^2 - k_\mathrm{x}^2,$ which does not yield any
corrections and also the factor $\exp{(  i  \pi(n-1)/4
)}\,(\pi/q)^{(n-1)/2}$ instead of such a factor for $n=2$.
Therefore, (\ref{gen-Four-2}) is modified as follows:
\begin{equation}
\widehat{\psi}(\vec{ k}) = A \, \left[k (k + k_1)^{\nu+(n-1)/2}
\right]^{-1}  \nonumber
\end{equation}
\begin{equation}
\times \exp\left[- i  k c t - (k - k_1)\frac{\varepsilon}{2} - (k +
k_1)\frac{\gamma}{2} - \frac{2 \kappa^2 \gamma}{ k + k_1} \right],
\label{gen-Four-many}
\end{equation}
where
\begin{equation}
A = (2\pi)^{n/2}\,\,  e ^{ i (n-1)\pi/4 } \,
\frac{p^{2\nu}}{\gamma^\nu}.
\end{equation}

\end{document}